\documentclass{PoS}
\usepackage{subfigure}
\usepackage{wrapfig}

\title{VERITAS Observations of The Galactic Center Ridge}

\ShortTitle{VERITAS Galactic Center Ridge}

\author{\speaker{Andrew W. Smith} for the VERITAS collaboration\thanks{veritas.sao.arizona.edu}\\
        University of Maryland, College Park / NASA Goddard Space Flight Center\\
        E-mail: \email{asmith44@umd.edu}}

\abstract{The Galactic Center Ridge has been observed extensively in the past by both GeV and TeV instruments revealing a wealth of structure, including both a diffuse component, the point sources G 0.9+0.1 (a composite supernova remnant) and SgrA* (believed to be associated with the super massive black hole located at the center of our galaxy). Previous observations ($>$ 300 GeV) with the H.E.S.S. array have also detected an extended TeV component along the Galactic plane due to either diffuse emission or a host of unresolved point sources. Here we report on the VERITAS observations of the Galactic Center Ridge from 2010-2014 in the energy range above 2 TeV. From these observations we 1.) Provide improved measurements of the differential energy spectra for SgrA* in the multi-TeV regime, 2.) Provide a detection in the $\>2$ TeV band of the composite SNR G 0.9+0.1 and an improvement of its multi-TeV energy spectrum. 3.) Report on the detection of an extended component of emission along the Galactic plane by VERITAS. 4.) Report on the detection of VER J1746-289, a localized enhancement of TeV emission along the Galactic plane.}

\FullConference{The 34th International Cosmic Ray Conference,\\
		30 July- 6 August, 2015\\
		The Hague, The Netherlands}

\begin{document}

\section{Introduction}

The Galactic Center Ridge (GCR) region is one of the most interesting local regions for study in the very high energy (VHE, $>$100 GeV) gamma-ray band. This is primarily due to its high concentration of star forming regions, pulsar wind nebulae (PWN), supernova remnants (SNR), and of course the central accelerator Sgr A*; all of which are known sources of VHE gamma rays.  Due to its high density of possible gamma-ray sources, the region has been studied extensively in the TeV band, \cite{WhippleGC, H.E.S.S.GC, MATTHIAS, MAGICGC}. The confirmed number of individual sources (point or extended) is relatively low for such a busy region ($<$ 5) with a very large proportion of gamma-ray emission in the region coming from either unresolved point sources, or a diffuse, extended component. Observations of this region with H.E.S.S. \cite{H.E.S.S.Diffuse} reveal a distinct band of emission stretching along the central region of the plane; this emission seemingly correlated with dense molecular cloud regions. As this diffuse component is assumed to be generated from cosmic ray interactions with the molecular clouds, the study of this region in the TeV band also allows for a characterization of the cosmic ray flux near the Galactic Center.
\begin{figure}[t]
\centering
\includegraphics[width=0.75\textwidth]{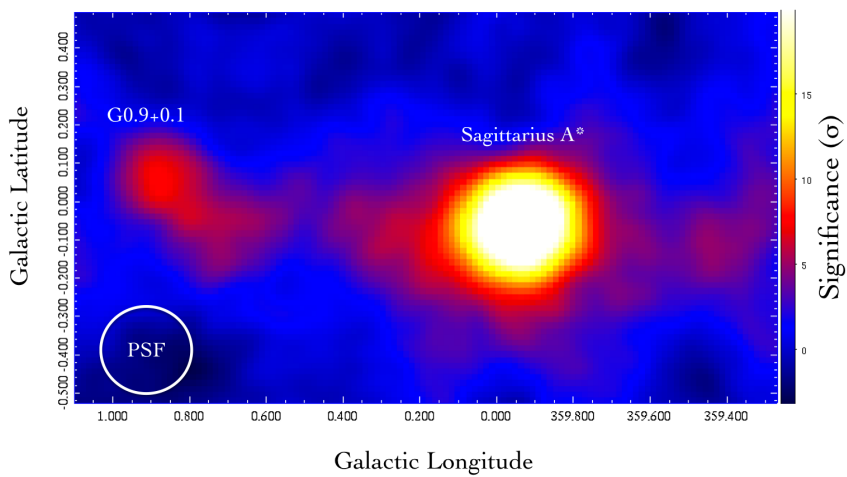}
\caption{The VERITAS $>$ 2 TeV significance map of the Galactic Center Ridge showing significant emission from Sgr A*, G0.9 +0.1, and a diffuse region between these two point sources.}
\label{fullmap}
\end{figure}
\section{VERITAS Observations}
The Very Energetic Radiation Imaging Telescope Array System (VERITAS), located at the Fred Lawrence Whipple Observatory (FLWO) in southern Arizona (31$^{\circ}$ 40' N, 110$^{\circ}$ 57' W,  1.3 km a.s.l.) is an array of four 12-meter imaging atmospheric Cherenkov telescopes (IACTs) providing excellent angular resolution and sensitivity to cosmic TeV gamma-ray sources \cite{HolderVTS}. In normal operations (i.e. high elevation observations), VERITAS is sensitive in the energy range of 85 GeV to $>$30 TeV and has the capability to detect a 1$\%$ Crab Nebula flux in less than 25 hours of observations. VERITAS has an energy resolution of 15$\%$ at 1 TeV and a typical angular resolution of $<$0.1$^{\circ} $.

\begin{figure}[t]
\centering
\includegraphics[width=1.0\textwidth]{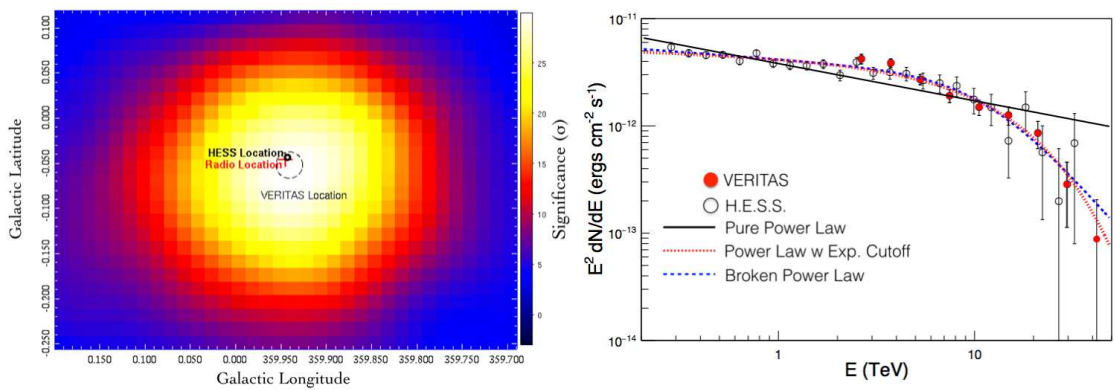}
\caption{Left: The VERITAS $>$ 2 TeV significance map of Sgr A* showing the VERITAS source location compared to the H.E.S.S. and radio locations. Right: The differential energy spectrum using both VERITAS and H.E.S.S. points along with the model fits described in the text.}
\label{fullmap}
\end{figure}

Between 2010-2014, VERITAS accrued $\sim$85 hours of quality selected, livetime observations of the Sgr A* region. Due to the Northern Hemisphere location of VERITAS, the Sgr A* region never transits above 30$^{\circ}$ elevation. This large zenith angle to the source results in a higher energy threshold (>2 TeV) for VERITAS observations. Normally, such observations would result in very poor angular resolution for ground based gamma-ray telescopes. To improve the performance,  a specialized analysis technique (see \cite{WhippleGC, MATTHIAS}) was used in which the displacement between the center of gravity of a parameterized Hillas ellipse and the location of the shower position within the camera plane is utilized. This displacement or "DISP'' method results in enhanced angular resolution at large zenith angle observations where small parallactic displacements between shower images would normally degrade angular resolution significantly.

These observations resulted in a significant detection of several distinct regions of $>$2 TeV emission in the GCR (significance skymap shown in Figure 1). The brightest source within the field is the central source Sgr A*, believed to be associated with the super massive black hole residing at the center of the galaxy. This data also allows for a $>$2 TeV detection of the composite supernova remnant G0.9 +0.1, as well as an extended  component of emission extending along the Galactic plane. In turn we examine all three of these detections and provide skymaps and spectra for both Sgr A* and G 0.9+0.1, and examine a new VERITAS source of TeV gamma rays located within the extended component along the plane.

\section{Sgr A*}
Previously identified as a VERITAS TeV source (VER J1745-290) \cite{MATTHIAS}), Sgr A* was detected at a  significance of 18$\sigma$ in approximately 46 hours of observations between 2010-2012. In the total of 85 hours of observations reported in this work, VERITAS detected a total of 735 excess gamma-ray events from VER J1745-290, resulting in a detection significance of $>$25$\sigma$. The resulting 2 dimensional significance map, is shown in Figure 2 along with both the H.E.S.S. ($>$300 GeV) \cite{H.E.S.S.GC} and radio locations \cite{radiolocation} of Sgr A*. The refined VERITAS position of VER J1745-290 is (Galactic coordinates) longitude = 359.94$^{\circ}$ $\pm$0.002$^{\circ}_{stat}$ $\pm$ 0.013$^{\circ}_{sys}$, latitude = -0.053$^{\circ}$ $\pm$0.002$^{\circ}_{stat}$ $\pm$ 0.013$^{\circ}_{sys}$, in good agreement with both of the radio and H.E.S.S. positions. 

The VERITAS differential energy spectrum from 2-30 TeV is shown, along with the H.E.S.S. spectral points in Figure 2. While the H.E.S.S. spectral points allow for accurate fits at lower energies, the VERITAS spectral points have significantly smaller errors at multi-TeV energies. By providing a joint fit to both the H.E.S.S. and VERITAS points from 0.2-50 TeV, the most accurate fit to the TeV emission can be obtained. Following the analysis of \cite{Aharonian et al 2009}, we investigated spectra following the shape of a.) a power-law, b.) a power-law with an exponential cutoff, and c.) a smoothly broken power-law. These fits have forms (respectively) of: 
\small
\begin{equation}
\frac{dN}{dE} = N_{0} \times \frac{E^{-\Gamma_{1}}}{1 \hspace{0.5mm}TeV}
\end{equation}

\begin{equation}
\frac{dN}{dE} = N_{0} \times \frac{E^{-\Gamma_{1}}}{1\hspace{0.5mm}TeV} \times e^{\frac{E}{E_{cut}}}
\end{equation}

\begin{equation}
\frac{dN}{dE} = N_{0} \times \frac{E^{-\Gamma_{1}}}{1\hspace{0.5mm}TeV} \times \frac{1}{1+(\frac{E}{E_{break}})^{\Gamma_{1}-\Gamma_{2}}}
\end{equation}
\normalsize
The fitting results (shown in Table 1, and in Figure 3) clearly exclude a power-law fit. While the exponential cutoff power-law and smoothly broken power-law models both provide adequate fits (reduced $\chi^{2}$ values close to 1.0), the fit for a smoothly broken power-law results in a normalization with a relatively large error. In the case of a power law with an exponential cut-off, the spectral parameters are in good agreement with previous measurements \cite{Aharonian et al 2009} and refine the location of the spectral cutoff to 12.8$\pm$1.9$_{stat}$ TeV. This cut-off parameter can used to provide more accurate constraints on energy of primary particles (leptons or hadrons) at work in the source. It should be noted that these spectral fits do not currently include any subtraction of the diffuse emission from the galactic plane (such as in \cite{Viana}), nor any estimate of the systematic error on the fit parameters.

\begin{table}[t]
\small
 \centering
 \begin{tabular}{c|c|c|c|c|c}
\textbf{Model}&  \textbf{N$_{0}$ ($\gamma$s cm$^{-2}$ s$^{-1}$ TeV$^{-1}$)} & \textbf{$\Gamma_{1}$} & \textbf{$\Gamma_{2}$} & \textbf{E$_{break}$/E$_{cut}$ (TeV)} & \textbf{$\frac{\chi^{2}}{n.d.f.}$}   \\\hline
Power Law &2.4 ($\pm$0.05) $\times$10$^{-12}$ & 2.35$\pm$0.02 & N/A & N/A &113/31\\\hline
Smoothly Broken & 3.62($\pm$3.77) $\times$10$^{-10}$ &4.14$\pm$0.37 & 2.1 $\pm$0.04 & 11.8$\pm$1.9 & 26.4/29 \\
Power Law & &&  &  \\\hline 
Exp. Cutoff  & 2.8($\pm$0.08) $\times$10$^{-12}$ & 2.1 $\pm$ 0.04 & N/A & 12.8 $\pm$ 1.9 & 28.18/30 \\
Power Law &  && &  & \\\hline
\end{tabular}
\caption{The results of the fitting of the VERITAS and H.E.S.S. spectral point of Sgr A* described in the text. }
\end{table}
\normalsize
\section{G 0.9+0.1}

The composite supernova remnant G0.9 +0.1 consists of a bright, compact radio PWN surrounded by an extended radio shell \cite{g0.9radio}. Estimated to have an age of a few thousand years \cite{Mezger, H.E.S.S.G0.9}, G 0.9+0.1 was first announced as a TeV source by the H.E.S.S. collaboration \cite{H.E.S.S.G0.9}, detecting $>$200 GeV emission at the level of approximately 2$\%$ of the Crab Nebula flux. The H.E.S.S. source is attributed to the PWN core of the remnant due to the morphology, as well as the apparent lack of hard X-ray emission in the shell remnant. The H.E.S.S. spectrum of the source from 0.2-7 TeV is well-fit by a simple power law with index of 2.3.

VERITAS observations of the GCR taken during 2010-2014 have also allowed for a statistically significant detection of G 0.9+0.1 in the $>$ 2 TeV regime. In the 85 hours of observations reported in this work, VERITAS detected a total of 134 excess events from G 0.9+0.1, corresponding to a statistically significant detection at the 7$\sigma$ level. The VERITAS source position (Figure 3) is centered at (Galactic coordinates) longitude = 0.86$^{\circ}$ $\pm$0.015$^{\circ}_{stat}$ $\pm$ 0.013$^{\circ}_{sys}$, latitude = 0.067$^{\circ}$ $\pm$0.02$^{\circ}_{stat}$ $\pm$ 0.013$^{\circ}_{sys}$ and  given the VERITAS source name VER J1747-281. The VERITAS position is coincident with the radio PWN, similar to the H.E.S.S. result. The VERITAS spectrum of G 0.9+0.1 (Figure 3) from 2-30 TeV is well fit (reduced $\chi^{2}$ of 3.1/9) by a power law (Eq. 3.1) with normalization (at 1 TeV) of 8.8 $\pm$ 0.8$_{stat}$ $\times$ 10$^{-13}$ photons TeV$^{-1}$ cm$^{-2}$ s$^{-1}$ and index of 2.3 $\pm$ 0.1$_{stat}$. We find no strong indications of a spectral break up to $\sim$22 TeV, placing a high limit on the parent population of relativistic electrons. This, along with the lack of a \textit{Fermi}-LAT detection of G 0.9+0.1 in the 0.3-100 GeV regime, may provide challenges to simple one zone models used to explain the emission from the source. A modeling of a the VERITAS+H.E.S.S. TeV spectral points, along with radio, x-ray and GeV data will be provided in an upcoming work.
\begin{figure}[t]
\centering
\includegraphics[width=1.0\textwidth]{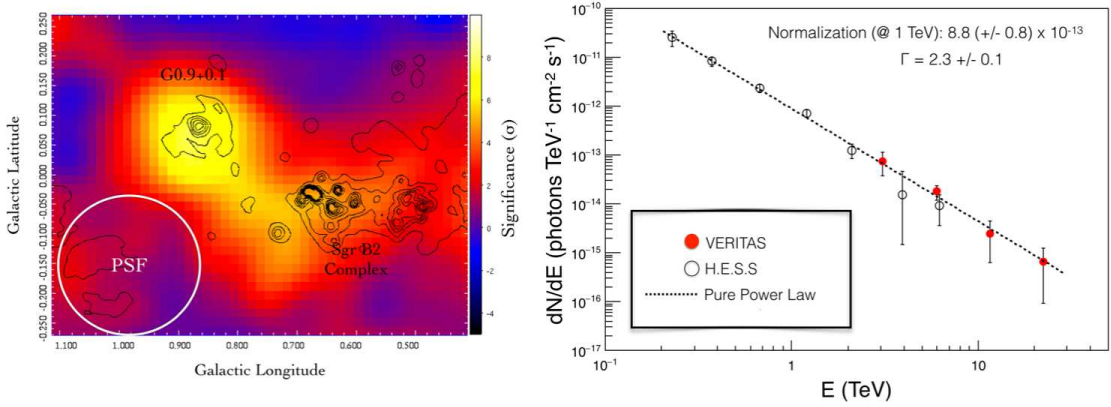}
\caption{Left: The VERITAS $>$ 2 TeV significance map of the composite SNR G0.9 +0.1, along with VLA 20cm radio contours \cite{Farhad}. The VERITAS location is consistent with the PWN of the SNR. The excess TeV emission from the Sgr B2 region can also be seen adjacent to G 0.9+0.1. Right: The differential energy spectrum of G 0.9+0.1 using both VERITAS and H.E.S.S. data points, the spectrum is well fit any a simple power law with no indication of a cutoff up to $\sim$20 TeV.}
\label{fullmap}
\end{figure}
\section{Ridge Emission and VER J1746-289}

In their discovery paper, the H.E.S.S. collaboration presented the residual maps (i.e. after subtracting known point sources within the field of view) of the $>$300 GeV emission from the Galactic Plane  \cite{H.E.S.S.Diffuse}. These residual maps revealed a complicated network of diffuse gamma-ray emission within the central 3$^{\circ}$ of the Galactic Plane. When plotted along with the CS emission contours, the H.E.S.S. emission appears correlated with dense molecular cloud regions (bright in CS line emission). However, given the complicated nature of the region, this measurement was unable to rule out the possibility of a significant contribution to the TeV flux coming from unresolved point sources. 
\begin{wrapfigure}{r}{0.5\textwidth}
  \begin{center}
    \includegraphics[width=0.6\textwidth]{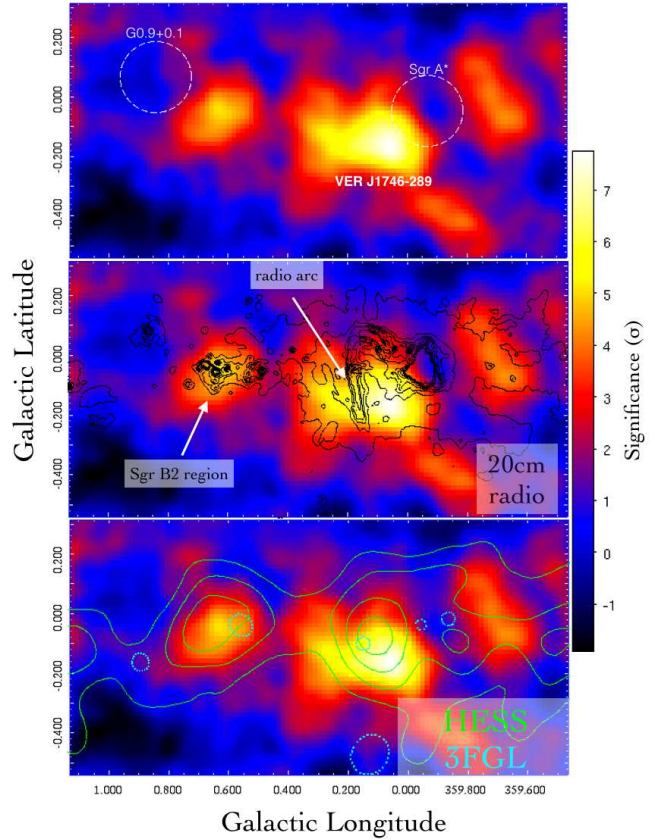}
  \end{center}
  \caption{The VERITAS $>$2 TeV signifiance maps of the GCR after subtracting excess emission from Sgr A* and G 0.9+0.1. The top panel shows the locations of the subtracted point sources as well as the VERITAS source VER J1746-289. VLA 20cm radio contours from {Farhad} are shown in the middle panel, with H.E.S.S. excess event contours and \textit{Fermi}-LAT 3FGL sources shown in the bottom panel.}
\end{wrapfigure}
To investigate whether the H.E.S.S. residual component is present in the VERITAS skymaps, we proceeded by removing the excess emission from the point sources (Sgr A* and G 0.9+0.1) within the field of view. The resulting significance skymap is shown in Figure 4 with radio (middle panel), and \textit{Fermi}-LAT/H.E.S.S. regions (bottom panel) plotted over the maps. VERITAS detects significant $>$2 TeV emission reaching $\sim1^{\circ}$ to the northeast (left in Figure 1) of Sgr A*, similar to the H.E.S.S. result \cite{H.E.S.S.GC}.  This extended component contains several local enhancements. For instance, we find an enhancement in TeV emission (similar to the H.E.S.S. result) co-located with the giant molecular cloud region Sgr B2 (see Figures 3 and 4). Additionally, we also find a statistically significant region of emission in the VERITAS skymaps adjacent to the location of Sgr A*.

The center of this excess is located at (Galactic Coordinates) longitude = 0.055$^{\circ}$ $\pm$0.01$^{\circ}_{stat}$ $\pm$ 0.013$^{\circ}_{sys}$, latitude = -0.148$^{\circ}$ $\pm$0.01$^{\circ}_{stat}$ $\pm$ 0.013$^{\circ}_{sys}$ and is given the name VER J1746-289. An extended source, VER J1746-289 is detected at a significance of 7.8$\sigma$ in the $\sim$85 hours of observations reported in this work.  There is a wealth of non-thermal emission structure that is consistent with the location of VER J1746-289. In radio, the famous Galactic radio arc can be seen adjacent to VER J1746-289 (Figure 4, middle panel). In the GeV/TeV regime, we see the H.E.S.S. excess contours and a \textit{Fermi}-LAT 3FGL source in Figure 4 (bottom panel).  Both of these GeV/TeV excesses provide a plausible association to VER J1746-289.

To further investigate possible multi-wavelength counterparts for VER J1746-289 we show the VERITAS (significance) and H.E.S.S. (excess) contours of the region in Figure 5 plotted over the XMM-Newton 2-7.2 keV count map. The known, non-thermal X-ray filament structure \cite{Johnson} is clearly apparent Figure 5, which overlaps well with the GeV/TeV emission in the region. Shown also in Figure 5 are the regions in this map identified to exhibit strong Fe K emission lines \cite{Ponti}, corresponding to dense molecular cloud regions. In \cite{Ponti}, it was shown that the X-ray emission (variable on a timescale of several years) from this region was primarily driven by superluminal shocks traveling through the molecular cloud region. It is unclear whether the shock originated from Sgr A* or from another nearby energetic source.

The morphology of VER J1746-289 is consistent with an association with both the non-thermal X-ray filaments shown in Figure 5 and the H.E.S.S. excess in the region. However, we also note that the peak significance of  VER J1746-289 appears to be leading away from both these sources (Figure 5). Further observations with VERITAS, compared to the latest results from H.E.S.S./H.E.S.S II and the MAGIC TeV observatories will help to resolve the nature of this apparent offset. If VER J1746-289 is indeed associated with the non-thermal X-ray filaments detailed in \cite{Ponti}, the study of the morphology of the TeV emission from this region can yield detailed information about the nature of particle acceleration in this region, as well as the density and distribution of molecular clouds within the inner region of the Galactic Center.

\begin{figure}[t]
\centering
\includegraphics[width=0.7\textwidth]{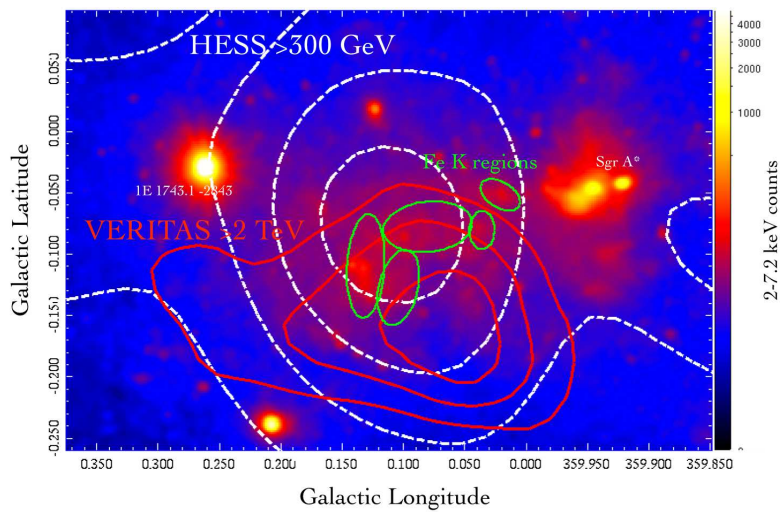}
\caption{The XMM-Newton hard X-ray (2-7.2 keV) map of the region adjacent to Sgr A*. The H.E.S.S. excess event contours (white) are shown along with the VERITAS 5,6,7 $\sigma$ significance contours. Also shown (green ellipses) are regions identified by Fe K emission to be particularly dense molecular cloud regions lit up by a superluminal shock.}
\label{fullmap}
\end{figure}

\section{Summary}

In this proceeding we present the contributions made by VERITAS to the study of the GCR and its myriad of structures. The VERITAS detections of both Sgr A* and G 0.9+0.1 above 2 TeV (in conjunction with the previous H.E.S.S. observations of these regions) allow for more accurate spectral measurements of these sources, providing key constraints to on going efforts to model the emission processes at work in both Sgr A* and G 0.9+0.1. In the case of Sgr A*, the more accurate constraint of the cutoff/break energy in the spectrum can lead directly to a better understanding of the accelerator responsible for the GeV/TeV emission. In the case of G 0.9+0.1, the lack of a break in the spectrum up to $\sim$20 TeV (as well as the lack of emission seen by \textit{Fermi}-LAT) places constraints on more simple emission models used to explain the emission from the PWN at the center of the remnant.

We also present "source subtracted" maps of the GCR, removing emission components from both G 0.9+0.1 and Sgr A* in order to reveal less prominent TeV gamma-ray emission features along the Galactic plane. This has revealed an extended structure extending along the plane to the east of Sgr A*, with several local enhancements evident in the maps. One of these enhancements, VER J1746-289,  is significantly detected by VERITAS and is most likely associated with already known non-thermal structures (radio, X-ray, GeV/TeV) in the region.

Further study of this region using these maps from VERITAS, as well as new maps from H.E.S.S/H.E.S.S. II, MAGIC, NuStar, and the VLA can yield a better understanding of the nature of the non-thermal sources in the Galactic Center Ridge.

\section{Acknowledgements}
This research is supported by grants from the U.S. Department of Energy Office of Science, the U.S. National Science Foundation and the Smithsonian Institution, and by NSERC in Canada. We acknowledge the excellent work of the technical support staff at the Fred Lawrence Whipple Observatory and at the collG 0.9+0.1aborating institutions in the construction and operation of the instrument. A.W. Smith acknowledges support through the Cycle 7 \textit{Fermi} Guest Investigator program, grant number NNH13ZDA001N.

 The VERITAS Collaboration is grateful to Trevor Weekes for his seminal contributions and leadership in the field of VHE gamma-ray astrophysics, which made this study possible.


\begin{thebibliography}{99}
\scriptsize
\bibitem{WhippleGC}
 Kosack, K.P., et al., \textit{"TeV Gamma-Ray Observations of the Galactic Center"} ApJ, 608, 97, (2004) \bibitem{H.E.S.S.GC}
Aharonian, F.  et al., \textit{"Very high energy gamma rays from the direction of Sagittarius A*"}, A$\&$A 425  pL13-L17  (2004)
\bibitem{MATTHIAS}
A. Archer et al. , \textit{"Very-high energy observations of the Galactic center region by VERITAS in 2010-2012"} ApJ, 790, 149 (2014)
\bibitem{MAGICGC}
J. Albert, et al., \textit{"Observation of Gamma Rays from the Galactic Center with the MAGIC Telescope"}, ApJ 638L.101A, (2006)
\bibitem{H.E.S.S.Diffuse}
Aharonian, F.A.,  et al., \textit{"Discovery of very-high-energy gamma rays from the Galactic Centre ridge"}, Nature, 439, 695, (2006)
\bibitem{HolderVTS}
Holder, J., et al., \textit{"Status of The VERITAS Observatory"}, AIP 303 Conf. Series, 1085, 657, (2008)
\bibitem{radiolocation}
Petrov, L. et al., \textit{"The VLBA Galactic Plane Survey -- VGaPS"}, Astron. J., 142, 35 (2011)
\bibitem{Aharonian et al 2009}
Aharonian, F. et al., \textit{"Spectrum and variability of the Galactic center VHE gamma-ray source HESS J1745-290"}, A$\&$A 503 p817-825 (2009)
\bibitem{Viana}
Viana, A., $\&$ Moulin, E., \textit{" Spectral analysis of the Galactic Center emission at very-high-energy gamma- rays with H.E.S.S"}, In Proc. of the ICRC2013, Rio de Janeiro (Brazil), (2013)
\bibitem{g0.9radio}
Helfand, D.J., Becker, R.H., \textit{"G0.9 + 0.1 and the emerging class of composite supernova remnants"}, ApJ., 314, 203., (1987)
\bibitem{Mezger}
Mezger, P.G., Duschl, W.J., Zylka, R.,  \textit{"Anatomy of the Sagittarius A complex. V. Interpretation of the SGR A* spectrum"}, A$\&$A Rv., 7, 289., (1996)
\bibitem{H.E.S.S.G0.9}
Aharonian, F.  et al. ,  \textit{"Very high energy gamma rays from the composite SNR G 0.9+0.1"}, A$\&$A 432, L25, (2005)
\bibitem{Farhad}
Yusef-Zadeh, F., Hewitt, J.W., and Cotton, W., \textit{"A 20 Centimeter Survey of the Galactic Center Region. I. Detection of Numerous Linear Filaments"},  ApJL, 155:421-550, (2004)
\bibitem{Johnson}
Johnson, S. P., Dong, H., Wang, Q. D, \textit{A large-scale survey of X-ray filaments in thCS e Galactic Centre}, MNRAS, 399, 3, (2009)

\bibitem{Ponti}
Ponti, G. et al., \textit{"Discovery of a Superluminal Fe K Echo at the Galactic Center: The Glorious Past of Sgr A* Preserved by Molecular Clouds"}, ApJ, 714, 732-747, (2010)

\end{thebibliography}
\end{document}